\documentclass[pra,amssymb]{revtex4}
\newcommand{\be}{\begin{equation}}
\newcommand{\ee}{\end{equation}}
\newcommand{\bea}{\begin{eqnarray}}
\newcommand{\eea}{\end{eqnarray}}
\usepackage{epsfig}
\usepackage{graphicx}

\begin{document}

\title{Rapidly Rotating Fermions in an Anisotropic Trap}

\author{N. Ghazanfari}
\email{gnader@newton.physics.metu.edu.tr} \affiliation{ Department
of Physics, Middle East Technical University, 06531 Ankara, Turkey }
\author{M.~\"O.~Oktel}
\email{oktel@fen.bilkent.edu.tr} \affiliation{ Department of
Physics, Bilkent University, 06800 Ankara, Turkey }

\date{\today}

\begin{abstract}
We consider a cold gas of non-interacting fermions in a two
dimensional harmonic trap with two different trapping frequencies
$\omega_x \leq \omega_y$, and discuss the effect of rotation on the
density profile. Depending on the rotation frequency $\Omega$ and
the trap anisotropy $\omega_y/\omega_x$, the density profile assumes
two qualitatively different shapes. For small anisotropy
($\omega_y/\omega_x \ll \sqrt{1+4 \Omega^2/\omega_x^2}$), the
density consists of elliptical plateaus of constant density,
corresponding to Landau levels and is well described by a two
dimensional local density approximation. For large anisotropy
($\omega_y/\omega_x \gg \sqrt{1+4 \Omega^2/\omega_x^2}$), the
density profile is Gaussian in the strong confining direction and
semicircular with prominent Friedel oscillations in the weak
direction. In this regime, a one dimensional local density
approximation is  well suited to describe the system. The crossover
between the two regimes is smooth where the step structure between
the Landau level edges turn into Friedel oscillations. Increasing
the temperature causes the step structure or the Friedel
oscillations to wash out leaving a Boltzmann gas density profile.

\end{abstract}

\maketitle

\section{Introduction}

Rotating ultracold gases have received a lot of attention after the
initial experimental demonstration of vortices and vortex lattices
in a rotating Bose Einstein
condensate\cite{Dalibard,Ketterle,Cornell}. Beyond vortex physics
these experiments hold promise to create clean, controllable
environments to investigate interesting quantum phases brought upon
by the unusually high effective magnetic fields created by rotation.
There are both theoretical proposals and experimental efforts to
investigate vortex lattice melting\cite{Cooper,Sinova}, Fractional
Quantum Hall states \cite{Baranov,Sorensen,Hafezi,Osterloh} and
BEC-BCS crossover for rotating
gases\cite{Greiner,Zwierlein,Zwierlein2,Bloch}.

The usual procedure for rotating gas experiments start by pumping
large amount of angular momentum into the cloud by various rotation
methods\cite{Ketterle, Cornell}. Later, confining the gas into trap
that is as isotropic as possible to preserve angular momentum, the
system settles into a ground state that is rotating at some
equilibrium frequency. Another alternative route, which is used in
at least one experiment\cite{Dalibard2}, is to create a rotating
anisotropic trap with some fixed rotation frequency and wait for the
system to reach the ground state at this fixed rotation frequency.
This procedure allows for one more control parameter, the trap
anisotropy, and is theoretically interesting as it allows the system
to smoothly go from a two dimensional regime for small anisotropy to
a one dimensional regime for large anisotropy.

The single particle Hamiltonian for a rotating anisotropic trap is
exactly diagonalizable, where the eigenenergies can be found by a
Bogoluibov transformation\cite{linn, Oktel}. However, a closed form
expression for the real space wavefunctions has not been given until
recently\cite{Fetter}. The investigations of rotating anisotropic
trap systems have focused either on the properties of vortices, or
dynamics in the extremely anisotropic
limit\cite{Sanchez,Sinha,Aftalion}. All of these studies concern the
behavior of interacting Bosons in an anisotropic rotating trap. The
interaction strength creates another energy scale which complicates
the crossover between the one dimensional and two dimensional
regimes of the anisotropic trap.

In this study, we consider the density profile for a gas of
non-interacting fermions in a rotating anisotropic trap. Because of
the Pauli suppression of s-wave scattering, spin polarized ultracold
fermi gases are naturally non-interacting. The absence of another
energy scale related to interactions simplifies the physics
considerably and reveals how the anisotropic trap smoothly connects
the one dimensional regime to two dimensional regime.

The density profile for non-interacting fermions are easily
calculated numerically by summing the densities of individual
eigenstates. For the rotating anisotropic trap the wavefunctions
consist of a exponential part multiplying a Hermite polynomial of a
complex coordinate\cite{Fetter}. The recursion relation for Hermite
polynomials allows for fast and accurate numerical calculation of
the density, enabling us to calculate the exact density for
thousands of particles for any rotation frequency or anisotropy.

The numerically calculated density profiles clearly show two
distinct regimes. For small anisotropy the density profile consists
of a ziggurat like structure of elliptical density plateaus. The
density inside  a plateau is an integer times $\frac{\gamma}{2 \pi}$
where $\gamma$ is a function of trap frequencies and rotation. Each
plateau is clearly identified to belong to a Landau Level, and this
regime can be understood as a small  deformation of the quantized
density profile in an isotropic rotating trap which was first
discusses by Ho and Ciobanu\cite{Ho}. As the trap anisotropy is
increased, the plateau structure is replaced by a profile that is
Gaussian in the strongly confined direction and a sum of semicircles
in the weakly confined direction. Friedel oscillations of the
density become prominent as this switching happens, and for very
large anisotropy the Landau Level quantization is replaced by
quantization of the wavefunction along the transverse, {\it i.e}
strongly confined, direction. The density profiles give a striking
example for smooth connection between the two dimensional and one
dimensional physics.

The paper is organized as follows: We review the single particle
energies and wavefunctions in the next section and give the
expressions for the wavefunctions in the lowest five Landau levels.
In section III we show that the crossover from two to one dimension
is already evident in the energy spectra and define the two regimes
quantitatively. Section IV focuses on the numerical density profiles
and the description of the properties in both regimes. We also
investigate the density profile analytically, and discuss one
dimensional and two dimensional local density approximation methods
along with Friedel oscillations. Section V contains the numerical
results at finite temperature. We conclude in section VI by briefly
discussing the experimental consequences of our results and
indicating directions for future investigations.

\section{Single Particle Problem}

We start by considering the single particle physics in a two
dimensional rotating anisotropic trap. In this section, we review
the results obtained by Fetter, who obtained expressions for the
wavefunctions in the Lowest Landau level in reference \cite{Fetter}.

Our discussion throughout the paper will be limited to behavior in
two dimensions. With this limitation, we are able to focus on the
physics of switching between one and two dimensional regimes without
the complications introduced by the motion in the third dimension.
Our results can be extended to three dimensional systems quite
easily if there is strong confinement in the third dimension or if
the potential in the third dimension is slowly varying\cite{Ho}.
Both of these limits are routinely realized in ultracold atom
experiments.

The Hamiltonian for a single particle in a two dimensional
anisotropic harmonic trap can be  written in the rotating frame as
\be \label{Hamiltonian} {\cal H} = \frac{1}{2 M} \left( p_x^2 +p_y^2
\right) + \frac{1}{2} M \omega_x^2 x^2+ \frac{1}{2} M \omega_y^2 y^2
- \Omega L_z. \ee Here $M$ is the mass of the particle, the angular
momentum in the $z$ direction $L_z$ is given as $L_z=x p_y - y p_x$,
and $\Omega$ is the rotation frequency. The trapping frequencies
along the $ x $ and $y$ directions are $\omega_x,\omega_y$,
respectively. Without loss of generality we take $\omega_y \geq
\omega_x$, and refer to the $y$ and $x$ directions as the strongly
confined and weakly confined directions throughout the paper. The
stability of the trapped system depends on the rotation frequency,
which requires the the rotation frequency to be smaller than the
weak confining frequency $\Omega \leq \omega_x$.

It is clear from the single particle Hamiltonian that the system is
essentially described by two dimensionless parameters, the
anisotropy \be \label{scaledwy}
\tilde{\omega}_y=\frac{\omega_y}{\omega_x} \geq 1, \ee and the
scaled rotation frequency \be \label{scaledOmg}
\tilde{\Omega}=\frac{\Omega}{\omega_x} \leq 1. \ee

All the relevant quantities can be non-dimensionalized by scaling
all the lengths by the oscillator length in the $x$ direction
$l_x=\sqrt{\hbar/m \omega_x}$, frequencies by $\omega_x$ and
energies by $\hbar \omega_x$. In the recent literature on rotating
anisotropic traps\cite{Aftalion,Oktel} different sets of
non-dimensional parameters have been used to describe the system,
unnecessarily complicating the interpretation of the results. We
avoid increasing this list of dimensionless quantities by following
the notation of Fetter\cite{Fetter} closely. Our numerical
calculations are done in dimensionless units, and we will explicitly
state our scaling where appropriate.

The eigenenergies for the Hamiltonian Eq.(\ref{Hamiltonian}) can be
obtained by a direct diagonalization, or two successive Bogoluibov
transformations\cite{linn,Oktel}. The eigenstates are labelled by
two integers $n \ge 0 $ and $m \ge 0$, and the corresponding energy
is \be \label{EnergyOfEigenstates} E_{n,m}= \hbar \left[
(n+\frac{1}{2}) \omega_- + (m+\frac{1}{2}) \omega_+ \right], \ee
where \be \omega_\pm^2=\omega_\perp^2+\Omega^2 \mp \sqrt{\frac{1}{4}
\left( \omega_y^2 - \omega_x^2 \right)^2 + 4 \omega_\perp^2
\Omega^2}, \ee with $\omega_\perp^2=\frac{1}{2} \left(
\omega_x^2+\omega_y^2 \right)$. A schematic plot of the energy
levels is given in figure 1, where one can observe that if the
rotation is rapid $ \Omega \approx \omega_x$, then $\omega_+ \ll
\omega_-$ and the levels with the same index $n$ are almost
degenerate. Using the analogy between a rotating system and a system
under a magnetic field, we identify the levels with the same $n$
index to belong to a Landau level, and call $n$ the Landau level
index.

While the system is controlled by the two dimensionless parameters
in Eqs. (\ref{scaledwy}), (\ref{scaledOmg}), a number of other
derived parameters give insight into the dynamics and allow one to
express results more concisely. The anisotropy of single particle
orbits are controlled by two dimensionless parameters \bea \beta_+
&=&
\frac{\omega_x^2 - \omega_+^2 - \Omega^2}{2 \Omega \omega_+},  \\
\nonumber \beta_- &=& \frac{\omega_-^2 - \omega_y^2 + \Omega^2}{2
\Omega \omega_+}, \eea both of which are between 0 and 1. We show in
the following sections that a particular combination of these
parameters \be \label{cParameter} c=\frac{1-\beta_+ \beta_-}{1 +
\beta_+ \beta_-}, \ee controls the switching between the one and two
dimensional regimes. The density of particles in a filled Landau
level (see Section IV) is related to the frequency \be
\label{gammaParameter} \gamma=\frac{\omega_-^2-\omega_+^2}{2
\Omega}. \ee Finally we define the two 'stretched' complex
coordinates which control motion in Landau levels \bea \xi_+ &=&
\sqrt{\frac{2 M
\gamma \beta_+}{\hbar}} \frac{x+i \beta_- y}{1+\beta_+ \beta_-}, \\
\nonumber \xi_- &=& \sqrt{\frac{2 M \gamma \beta_-}{\hbar}}
\frac{y+i \beta_+ x}{1+\beta_+ \beta_-}. \eea

The ground state of the system has the wavefunction \bea
\label{GroundState} \phi_{00}=\frac{1}{\sqrt{\pi a_x a_y}}
\exp{\left[-\frac{x^2}{2 a_x^2} - \frac{y^2}{2 a_y^2} \right]}
\exp{\bigg\{i \frac{M x y}{\hbar} \left[\frac{\gamma}{1+\beta_+
\beta_-} - \frac{1}{2} \left(\frac{\omega_+}{\beta_+} +
\frac{\omega_-}{\beta_-} \right) \right] }\bigg\}, \eea where $a_x$
and $a_y$ are the widths of the Gaussian envelope \be
a_x^2=\frac{1+\beta_+ \beta_-}{\beta_+} \frac{\hbar}{M \gamma},
\hspace{.1cm} a_y^2=\frac{1+\beta_+ \beta_-}{\beta_-} \frac{\hbar}{M
\gamma}. \ee

The wavefunctions in the lowest Landau level (LLL) are found by
applying the relevant raising operator to be \be
\label{LLLwavefunctions} \phi_{m0}(x,y)= \frac{1}{\sqrt{m!}} \left(
\frac{c}{2} \right)^{m/2} H_m\left(\frac{\xi_+}{\sqrt{2 c}}\right)
\phi_{00}\left(x,y\right), \ee where $H_m$ is the $m^{th}$ Hermite
polynomial.

Another raising operator allows one to generate the wavefunctions in
the higher Landau Levels, for example the states in the first
excited Landau level have the wavefunction \bea
\label{1LLwavefunctions} \phi_{m1}(x,y)= \frac{1}{\sqrt{m!}} \left(
\frac{c}{2} \right)^{(m-1)/2}\phi_{00}\left(x,y\right)\left[ \xi_-
\frac{c}{2} H_m\left(\frac{\xi_+}{\sqrt{2 c}}\right) -2 i m \rho
H_{m-1}\left(\frac{\xi_+}{\sqrt{2 c}}\right) \right], \eea with
$\rho=\sqrt{\beta_- \beta_+}/(1+\beta_+ \beta_-)$. We give the
wavefunctions for the six lowest Landau levels in the appendix, and
use them in the numerical calculations in section III.

\begin{figure}
\begin{center}
\includegraphics[width=4in]{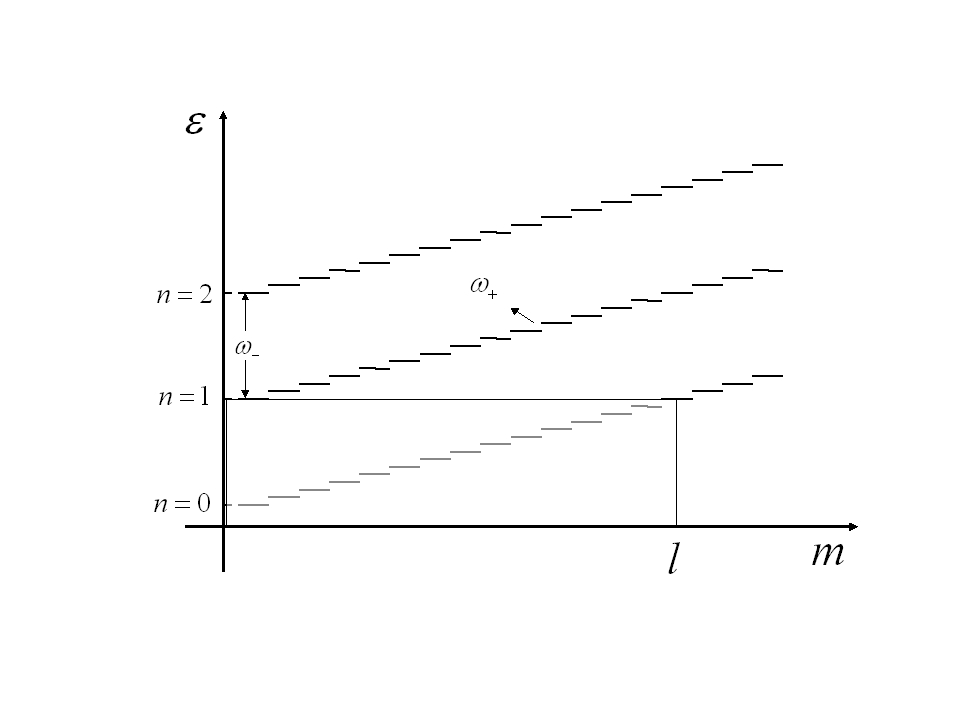}
\end{center} \caption{Energy vs. number of states in each Landau
levels. \emph{l} is the highest number of states within LL's before
the next LL is excited.} \label{fig1}
\end{figure}

The wavefunctions given above are remarkable, as they continuously
connect the usual 1D harmonic oscillator wavefunctions which have
the Hermite polynomial form to 2D Landau level wavefunctions which
are (for the lowest Landau level) analytic functions of the complex
coordinate $z=x+iy$. In the isotropic limit the $m^{th}$
wavefunction has a $m^{th}$ order zero at the origin. As the
anisotropy is turned on the $m^{th}$ order zero immediately breaks
up into $m$ first order zeros all of which are on the $x$ axis, as
the roots of Hermite polynomials are always real\cite{Oktel}.
Increasing anisotropy separates the roots from each other and forces
the Gaussian envelope to be more and more anisotropic. In the next
section we analyze the one and two dimensional regimes by
concentrating on a single quantity that is found from the
eigenvalues: the number of energy states in the LLL that have less
energy than the first excited Landau Level.

\section{One and two dimensional regimes}

The switching between the two dimensional and one dimensional
regimes is already apparent in the energy spectrum given in
Eq.(\ref{EnergyOfEigenstates}). It is institutive to consider two
limiting cases. The energies of eigenstates of a rotating isotropic
trap given as \be E_{nm}=\hbar \left[ \left( n+\frac{1}{2} \right)
(\omega + \Omega) + \left( m+\frac{1}{2} \right) (\omega-\Omega)
\right]. \ee The energies for a non--rotating anisotropic trap are
\be E_{n_x n_y} = \hbar \left[ (n_x + \frac{1}{2} ) \omega_x + (n_y
+ \frac{1}{2} ) \omega_y \right]. \ee Both of these expressions have
the same structure, as they are labeled by two non-negative
integers, and this structure can be graphically represented as in
figure 1. The rotating anisotropic trap system interpolates between
these two limits, and a good identifier for this interpolation is
the number of lowest Landau Level states that have lower energy than
any state in the higher Landau levels. We define \be \label{defineL}
l=\frac{\omega_-}{\omega_+}, \ee which is graphically shown in
figure 1.

\begin{figure}
\label{fig2}
\begin{center}
\includegraphics[width=3.5in]{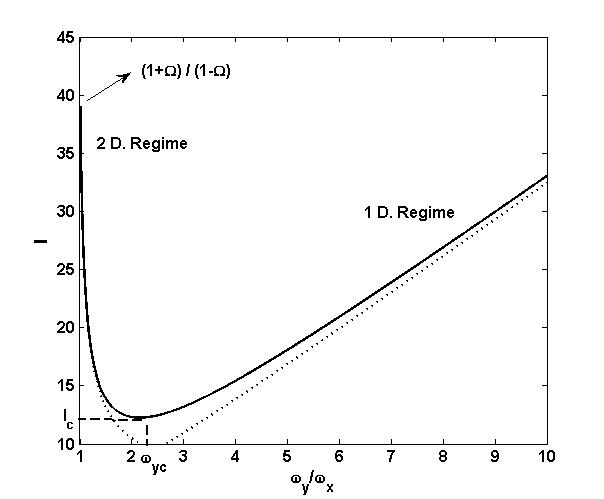}
\end{center} \caption{Solid line shows the behavior of l, number of the stats within two LL's as a
function of the anisotropy of the system $\omega_y/\Omega_x$. Dashed
lines are the asymptotic results for one and two dimensional limits.
Here $\omega_{yc}=(1+4\Omega/\omega_x)^{1/2}$ is the critical
anisotropy where l is minimum. The minimum number of states is given
by $l_c=(\frac{8}{1-\Omega/\omega_x})^{1/2}$}
\end{figure}

The behavior of $l$ as a function of $\omega_y/\omega_x$ at fixed
rotation frequency $\Omega/\omega_x$ clearly shows the two regimes
(See Fig.(2)). At zero anisotropy, $l$ has its isotropic value
$(1+\tilde{\Omega})/(1-\tilde{\Omega})$. When the anisotropy is
introduced, this value quickly decreases as
$(\tilde{\omega}_y-1)^{-1/2}$ and obtains its minimum value of
$l_{\rm{min}}=\sqrt{8/(1-\tilde{\Omega})}$ at
$\tilde{\omega}_y=\sqrt{1+4 \tilde{\Omega}^2}$. As the anisotropy is
increased further, $l$ increases linearly with $\tilde{\omega}_y$
with slope $1/\sqrt{1-\tilde{\Omega}^2}$. This linear increase is
exactly what is expected in the one dimensional regime as here $l$
would be the ratio of the effective frequencies in the $y$ and $x$
directions, \be
l=\sqrt{\frac{\omega_y^2-\Omega^2}{\omega_x^2-\Omega^2}}\simeq
\frac{\tilde{\omega}_y}{\sqrt{1-\tilde{\Omega}^2}}.\ee

The degeneracy of the Landau levels in the two regimes originate
from two different physical effects. When the anisotropy is small,
and the system is in the two dimensional regime, Coriolis force
induced by the rotation partially freezes the kinetic energy of the
particles and causes the high degeneracy within the Landau levels.
For high anisotropy one can think of the energy spectrum in terms of
subband quantization in a narrow channel, where each 'Landau Level'
actually is formed by states that have the same wavefunction in the
tightly confined direction.

The form of the wavefunctions also support this identification based
on $l$. For low anisotropy the wavefunctions peak around an ellipse
in the $x-y$ plane, while in the one dimensional regime the density
in the $x$ direction shows prominent oscillations, similar to a one
dimensional harmonic oscillator wavefunctions. In the next section,
we show that the density of a Fermi gas trapped in a rotating
anisotropic trap has markedly different behavior in these two
regimes.

\section{Density Profiles}

We now consider the density of a gas of $N$ non-interacting spinless
fermions in a rotating anisotropic trap. As the real space
wavefunctions are known the calculation of density reduces to a sum
over the absolute squares of the wavefunctions of all filled
eigenstates. Hence the density is given by \be \label{density}
\rho(x,y)= \sum_{nm} \left| \phi_{nm}(x,y) \right|^2
\theta(\mu-E_{nm}), \ee where $\mu$ is the chemical potential and
the energies $E_{nm}$ and corresponding wavefunctions $\phi_{nm}$
are given in the previous sections. The chemical potential
determined by the number of particles $N$ in the system with the
constraint \be \int dx dy \rho(x,y) = N. \ee

The LLL wavefunctions given in sec II are composed of an exponential
factor multiplying a Hermite polynomial, and wavunctions in the
first few Landau levels given in the appendix also have a similar
structure. Hermite polynomials, like other orthogonal polynomial
satisfy a recursion relation\cite{Arfken} \be H_{n+1}(x) = 2 x
H_{n}(x) - 2 n H_{n-1}(x), \ee which enables rapid and accurate
numerical evaluation of the wavefunctions. Because of the recursive
structure, the density sum Eq.(\ref{density}) can also be calculated
numerically at very low computational cost. We calculated density
profiles for up to $N=2000$ particles, for parameter values such
that all the particles reside in the lowest six Landau Levels.

\begin{figure}
\begin{tabular}{cc}
\includegraphics[scale=0.4]{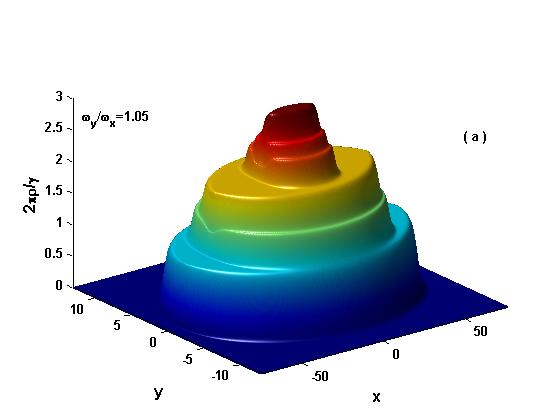}&
\includegraphics[scale=0.4]{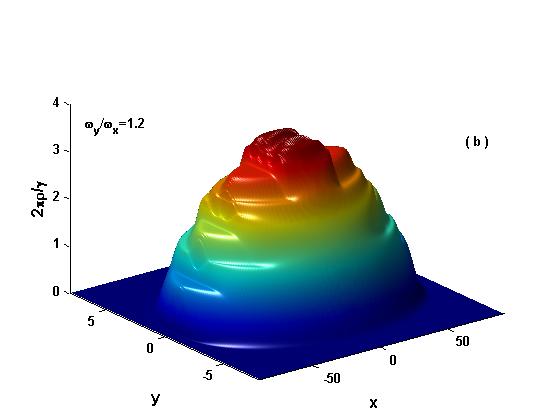}
\end{tabular}
\begin{tabular}{cc}
\includegraphics[scale=0.4]{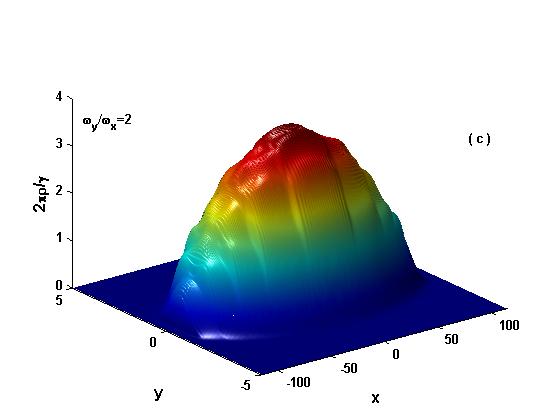} &
\includegraphics[scale=0.4]{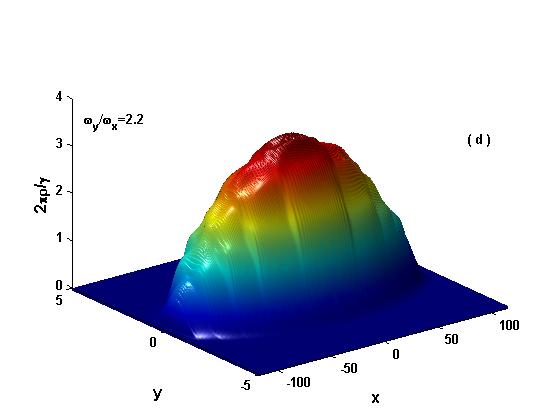}
\end{tabular}
\begin{tabular}{cc}
\includegraphics[scale=0.4]{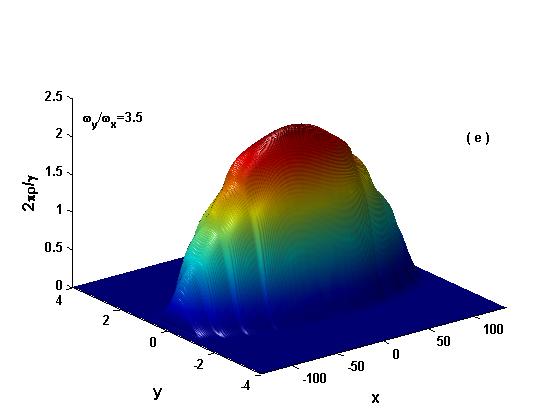} &
\includegraphics[scale=0.4]{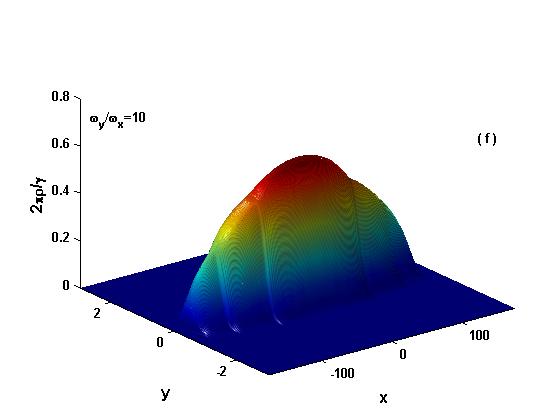}
\end{tabular}
\caption{(color online) Density profiles for fermions at different
anisotropy values of the system and fixed number of the particles
N=1000, and fixed rotation frequency $\Omega/\omega_x=0.999$. The
Friedel oscillations are observed in density profiles of anisotropic
cases. The number of LL's that fermions fill at each anisotropy can
also be determined by counting the number of plateaus in the density
profiles.}
\end{figure}

In figure 3, we display the density profiles for $N=1000$ particles
with fixed rotation frequency $\Omega/\omega_x=0.999$, for different
values of anisotropy. These parameters demonstrate the general
behavior of the density profile when the particles are distributed
over the first few Landau levels. For small anisotropy the density
profile mainly consists of steps, with small regions of switching
between them. Each plateau is clearly linked with a Landau level,
and defines an elliptical area of almost constant density. As shown
below, the density contribution from a filled Landau level is
$\rho=\frac{\gamma}{2 \pi}$. Thus, the steps have density equal to
an integer times $\frac{\gamma}{2 \pi}$. For the parameters in
Fig.(3a) only the lowest three Landau levels have particles and
corresponding steps can be identified clearly. The switching between
the $n^{th}$ step and $(n-1)^{th}$ steps happens with $n-1$
oscillations in density, which can also be seen in the figure. As
the anisotropy is increased the density oscillations between the
Landau level steps become more prominent especially in the weak
trapping direction. When the anisotropy is within the critical
region of switching between two and one dimensional behavior, the
steps and the Landau level structure become smeared out. Instead,
one can observe the density oscillations becoming prominent
throughout the cloud. These oscillations are expected for Fermions,
as a sharp Fermi surface cutoff in momentum space results in
oscillations in real space, known as Friedel
oscillations\cite{Freidel}. For larger values of anisotropy, the
density profile assumes a Gaussian shape in the strong confining
direction and the familiar semicircular shape of one dimensional
trapped fermions in the weak confining direction. Each Landau Level,
or rather sub-band in this regime, adds a new semicircle in this
direction, thus it is quite easy to observe how many sub-bands are
occupied form the density profile (See Fig.(3f)).

\begin{figure}
\begin{tabular}{cc}
\includegraphics[scale=0.4]{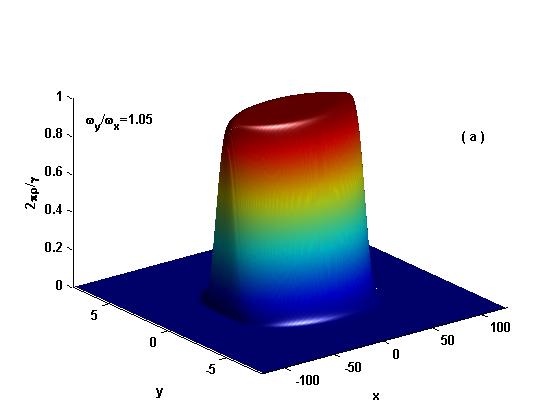}&
\includegraphics[scale=0.4]{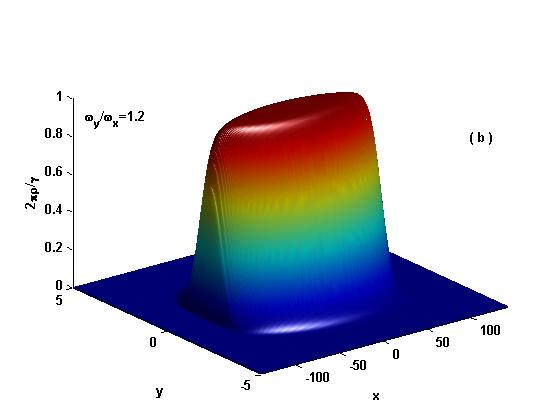}
\end{tabular}
\begin{tabular}{cc}
\includegraphics[scale=0.4]{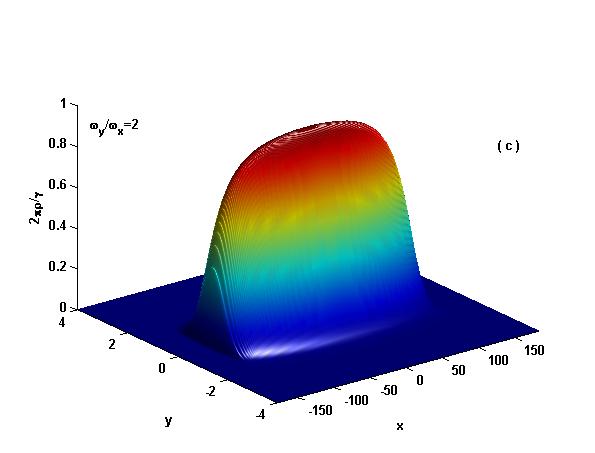} &
\includegraphics[scale=0.4]{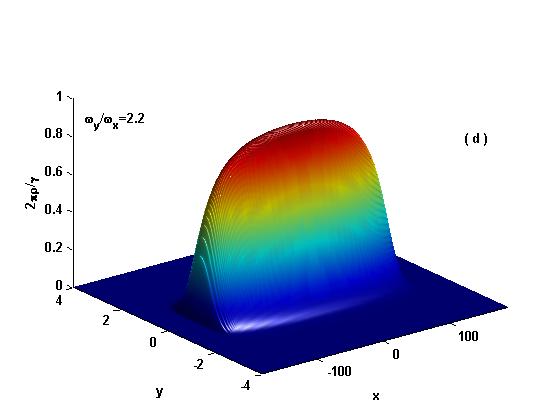}
\end{tabular}
\begin{tabular}{cc}
\includegraphics[scale=0.4]{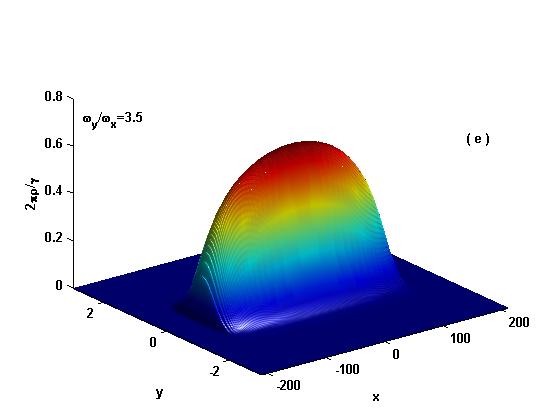} &
\includegraphics[scale=0.4]{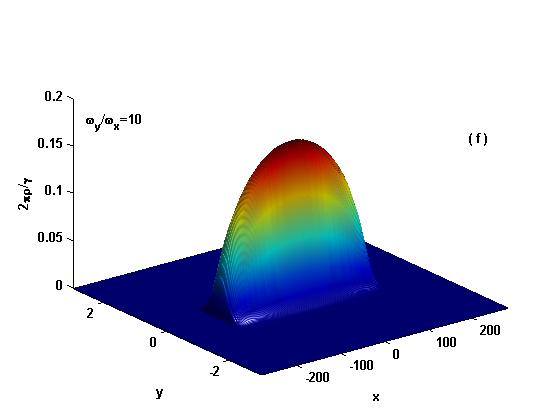}
\end{tabular}
\caption{(color online) Density profiles of fermions at different
anisotropy values of the system and fixed number of the particles
N=200, and fixed rotation parameter $\Omega/\omega_xa=0.9999$, when
all the fermions are settled at the LLL. The Friedel oscillations
are also observed for density profiles of anisotropic cases.}
\end{figure}

 In figure 4 we display the density profiles with N=200,
a low enough particle number so that all the particles remain in the
LLL for any value of anisotropy. In this figure, it is easier to
discern the character of the change between the two and one
dimensional regimes, as one can observe how the flat plateau is
replaced by a semicircular shape with Friedel oscillations. Again,
the switching happens around the critical value of anisotropy
determined from $l$.

To gain a better understanding of the density profile, we
investigate the density sum Eq.(\ref{density}) in more detail. Let's
concentrate on the case for which all the atoms reside in the LLL,
for  N particles we can write the sum as \bea \rho(x,y)&=&
\sum_{m=0}^N \left| \phi_{0m}(x,y) \right|^2 \\ \nonumber &=&
\sum_{m=0}^N \frac{1}{m!} \left( \frac{c}{2} \right)^m
H_m(\frac{\bar{\xi}}{\sqrt{2c}})
H_m(\frac{\xi}{\sqrt{2c}})\frac{1}{\pi a_x a_y}
e^{-x^2/a_x^2-y^2/a_y^2}. \eea In the limit $N \rightarrow \infty$
this sum can be evaluated using the generating function for Hermite
polynomials. We define \be S(\bar{z},z)= \sum_{m=0}^\infty
\frac{1}{m!} \left( \frac{c}{2} \right)^m H_m(\bar{z}) H_m(z). \ee
We use the generating function for Hermite polynomials \be
e^{-t^2+2tz}=\sum_{m} \frac{H_m(z)}{m!} t^m, \ee but regard $t = |t|
e^{i \theta}$ as a complex variable. Then the infinite sum can be
written as \bea S(\bar{z},z)= \int_0^{2 \pi} \frac{d \theta}{2 \pi}
\int_0^\infty d|t| 2 |t|\exp{\left[ -\frac{c}{2}(\bar{t}^2+t^2) +
\sqrt{2 c} (\bar{t} \bar{z} + t z) - |t|^2\right]}. \eea The
resulting Gaussian integral can be evaluated to yield \bea
S(\bar{z},z)=\frac{1}{\sqrt{(1-c)(1+c)}} \exp{\bigg[\frac{1}{2}
\frac{c}{1-c} (i (z-\bar{z}))^2\bigg]}\exp{\bigg[\frac{1}{2}
\frac{c}{1+c} (z+\bar{z})^2\bigg]}, \eea which gives the constant
density \be \rho(x,y)=\rho_0=\frac{1}{\sqrt{1-c^2}} \frac{1}{\pi a_x
a_y}= \frac{m \gamma}{2 \pi \hbar}. \ee We see that $\gamma$ is the
parameter that controls the density of states per Landau level in an
anisotropic trap. In the numerical calculations we see that the
steps in the density profile appear at integer multiples of this
value.  In the two dimensional regime, when the anisotropy is small,
the density at the center of the trap quickly reaches this value,
even for small number of particles. This is because the contribution
of higher wavefunctions to the density near the center is small for
low anisotropy.

We must remark that the above result is correct for any anisotropy,
i.e. if one puts infinitely many particles into the LLL one would
get a constant density. However as the particle number is increased
the cloud goes through two very different paths to this constant
density profile. In the two dimensional regime, the sum of the
density of $N$ particles defines an elliptical  region of constant
density  which grows in size as $N$ is increased. In the one
dimensional regime, density is very anisotropic and shows prominent
Friedel oscillations. The frequency of these oscillations decreases
as $N$ is increased and the density becomes smooth very slowly with
$\sim 1/N$.

The presence of Friedel oscillations in a trapped Fermion system is
not unexpected, as the physical reason for their presence is the
sharp cutoff at Fermi energy. When one is in the two dimensional
Landau Level regime the oscillations are suppressed as locally the
maximum allowed density within a Landau level is reached. One can
investigate the Friedel oscillations by a simple scaling argument as
follows. First let us define the finite sum over Hermite polynomials
as \be S_N(a,b,c)= \sum_{n=0}^N \frac{H_n(\bar{z})H_n(z)}{2^n n!}
c^n, \ee with $z=a+ib$. Now because of the infinite sum result we
have obtained above we expect two such sums to be related as \bea
S_N(a,b,c_2)\cong \sqrt{\frac{1-c_1^2}{1-c_2^2}}
S_N\bigg(\sqrt{\frac{(1+c_1)c_2}{(1+c_2)c_1}}a,
\sqrt{\frac{(1-c_1)c_2}{(1-c_2)c_1}}b,c_1\bigg), \eea if $N \gg 1$
is sufficiently large. Now $c_1$ can be chosen arbitrarily close to
one, \bea S_N(a,b,c_1)&=&\sum_{n=0}^N
\frac{H_n(\bar{z})H_n(z)}{2^n n!} c_1^n \nonumber\\
&\approx& \sum_{n=0}^N \frac{H_n(\bar{z})H_n(z)}{2^n n!}\nonumber\\
&=& S_N(a,b,1), \eea provided that $(1-c_1)N \ll 1$. This last sum
can be evaluated exactly using Christoffel-Darboux
formula\cite{Abramowitz}, \bea S_N(a,b,1)&=&\sum_{n=0}^N
\frac{H_n(\bar{z})H_n(z)}{2^n n!}\nonumber\\ &=& \frac{
H_{N+1}(\bar{z}) H_N(z) - H_{N+1}(z) H_N(\bar{z})}{2^{N+1} N!
(\bar{z}-z)}. \eea This exact evaluation and the scaling assumption
readily yields a formula for the wavevector of the Friedel
oscillations near the center of the trap \be \rho(x,y=0) = \rho_0
\left( 1 + (-1)^N \frac{1}{4 N} cos(k_F x) \right), \ee with \be k_F
l_x = \sqrt{8N/\omega_x} \sqrt{\omega_+ + \omega_- \beta_+ \beta-}.
\ee As expected, this scaling result overestimates the amplitude of
the Friedel oscillations, especially in the two dimensional regime.
The suppression of oscillations due to limited local density of
states is not captured by the scaling assumption. However, scaling
correctly describes the wavelength of the oscillations well into the
regime where they are too small to observe within our numerical
precision.

\begin{figure}
\includegraphics[width=3.5in]{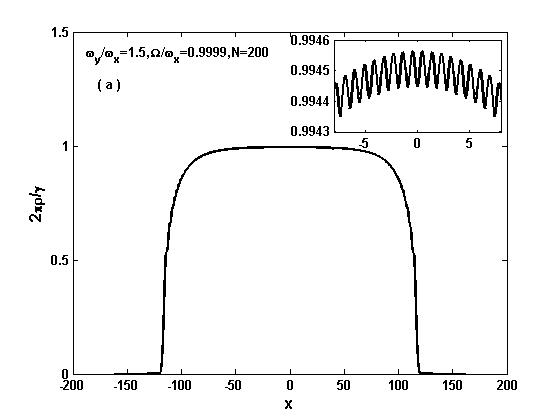}
\includegraphics[width=3.5in]{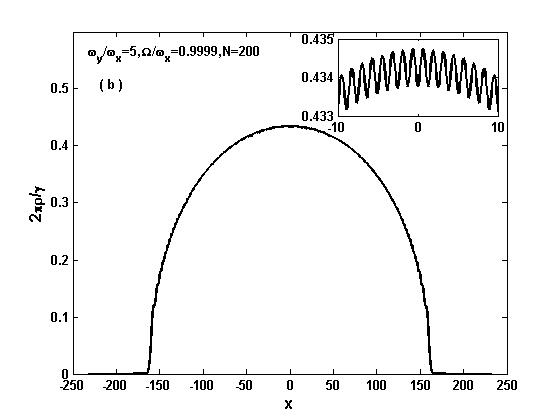}
\caption{(a) Density for N=200 fermions at $\Omega/\omega_x=0.9999$,
and $\omega_y/\Omega_x=2.2$. (b) Density for N=300 fermions at
$\Omega/\omega_x=0.999$, and $\omega_y/\Omega_x=2.2$. The insets in
both figures highlight the Friedel oscillation in density profiles
of the fermions.}
\end{figure}

When the external potential is smooth, a very useful tool to
describe the density profile is local density approximation. For the
case of a rotating anisotropic lattice we need two different
versions of LDA to describe the different behavior in the two
regimes.

For the two dimensional regime the effective potential in both
strong and weak confining directions are smooth, when compared with
the magnetic length imposed by rotation. Thus, it becomes
permissible to treat the system locally as a homogenous system with
the Landau Level structure. The local density at any point of the
gas will be an integer multiple of the density of a filled Landau
level $m \Omega / \hbar$. Thus the density profile within the 2D LDA
consists of a sequence of elliptical plateaus, which matches well
with the numerically calculated density profiles in the 2D regime.
Thus 2D LDA is successful in describing the density profile except
for the switching regions between the Landau level steps. A typical
plot comparing the result of LDA with the calculated density profile
is given in Fig.(6).

\begin{figure}
\includegraphics[width=3.5in]{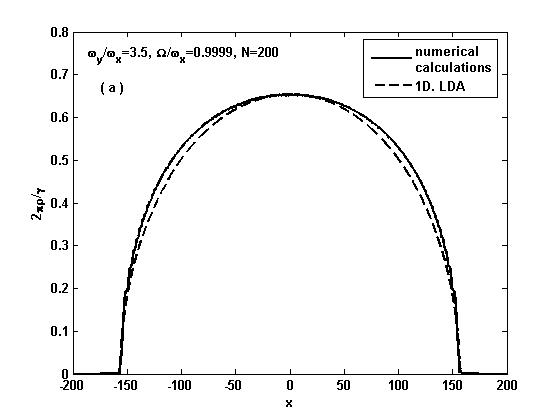}
\includegraphics[width=3.5in]{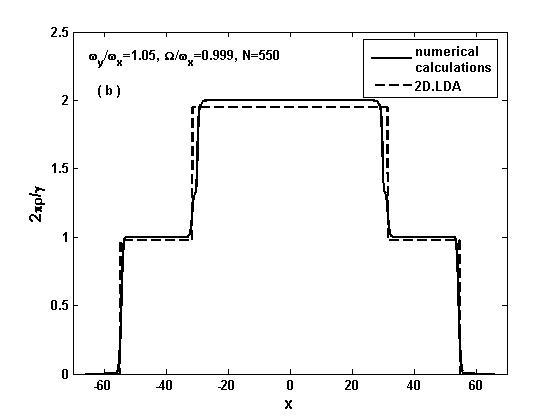}
\caption{(a) Density of N=200 fermions obtained by direct numerical
calculations (solid line), and one dimensional LDA (dashed line) at
$\Omega/\omega_x=0.9999$, and $\omega_y/\Omega_x=3.5$. (b) Density
of N=325 fermions obtained by direct numerical calculations (solid
line), and two dimensional LDA (dashed line) at
$\Omega/\omega_x=0.99$, and $\omega_y/\Omega_x=1.2$.}
\end{figure}

As the anisotropy is increased and the gas enters the crossover
regime the step structure in the density profile is no longer
observed. This regime is not well described by a two dimensional
LDA, as the potential in the strong confining direction is no longer
smooth on the scale of magnetic length. However if the anisotropy is
increased further, the oscillator length in the strong confining
direction becomes much smaller than the magnetic length and another
LDA approach becomes feasible. This is exactly the one dimensional
regime described in Sec. III.

In the one dimensional regime, the density profile in the strong
confining direction is almost the same for all wavefunctions in the
same Landau level. Thus we can treat the system within a one
dimensional LDA and obtain the usual semicircular profile for the
one dimensional fermions \be \rho(x,y=0)= \rho(0,0) \sqrt{1-
\frac{x^2}{L^2}}, \ee where the radius of the cloud is found as \be
L/l_x = \sqrt{\frac{2 \mu-\omega_-/\omega_x}{1-\tilde{\Omega}^2}}.
\ee As can be observed in Fig.(6), this approximation describes the
density profile in the one dimensional regime reasonably. The
agreement increases with increasing anisotropy, but Friedel
oscillations can not be captured within the local density
approximation. The agreement of one dimensional LDA with the
numerically calculated profiles further support our interpretation
of the very anisotropic Landau levels as sub-bands formed by
quantization in the strong confinement direction.

\section{Temperature effects}

The density profiles given in the previous sections were all
calculated at zero temperature. At finite temperature the density
profile can be calculated by including the fermi distribution
function as \be \label{temperaturedensity} \rho(x,y)= \sum_{nm}
\left| \phi_{nm}(x,y) \right|^2 \frac{1}{\exp{\beta(E_{nm}-\mu)}+1},
\ee where $\beta=\frac{1}{k_B T}$ is the inverse temperature.

We calculated density profiles for finite temperatures using the
same numerical procedure with the zero temperature case. The
temperatures and chemical potentials were chosen to ensure that the
occupation of the states higher than the sixth Landau level is less
than $exp[-5]\sim 0.0067$. For the two dimensional regime the step
structure of the density profile is expected to be smeared out when
the temperature is enough to close the gap between the Landau levels
$k_B T \sim \hbar \omega_-$. The calculated density profiles show
that the step structure becomes indiscernable below this value, as
the transition regions between the Landau level steps dominate the
profile. For larger temperatures the profile assumes the gaussian
profile as would be expected from a Boltzmann gas.

For the one dimensional regime, the first effect of the finite
temperature is to smear out the Friedel oscillations. The relevant
energy scale for Friedel oscillations is $\hbar \omega_+$ the
separation of energy levels within a Landau level, and as soon as
this temperature is reached the density profile becomes smooth,
while still obeying the semicircular LDA result. For higher
temperatures, once again the density profile becomes Gaussian as
would be expected from a Boltzmann gas. The effects of temperature
in both regimes is plotted for typical density profiles in figure 7.

\begin{figure}
\includegraphics[width=3.5in]{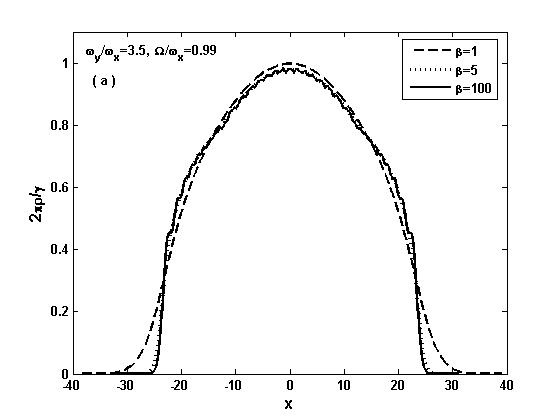}
\includegraphics[width=3.5in]{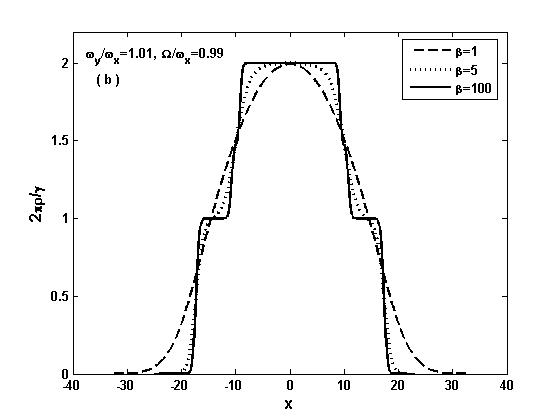}
\caption{(a) Density profile for fermions at different temperatures
and at $\Omega/\omega_x=0.99$, and $\omega_y/\Omega_x=3.5$. (b)
Density profile for fermions at different temperatures and at
$\Omega/\omega_x=0.99$, and $\omega_y/\Omega_x=1.01$. The
oscillations (a) and the layered aspect of the profiles (b) are
eliminated by temperature. Here $\beta=\frac{\hbar\omega_x}{k_bT}$
is the normalized inverse temperature. }
\end{figure}

\section{Conclusion}

We show that the density profile of a non-interacting Fermi gas in a
rotating anisotropic trap has two distinct regimes depending on the
rotation rate and anisotropy. Through numerical calculation and a
local density approximation the distinct behavior in the one and two
dimensional regimes is observed. For small anisotropy
($\omega_y/\omega_x \ll \sqrt{1+4 \Omega^2/\omega_x^2}$), the
density consists of elliptical plateaus of constant density,
corresponding to Landau levels and is well described by a two
dimensional local density approximation. For large anisotropy
($\omega_y/\omega_x \gg \sqrt{1+4 \Omega^2/\omega_x^2}$), the
density profile is Gaussian in the strong confining direction and
semicircular with prominent Friedel oscillations in the weak
direction. In this regime, a one dimensional local density
approximation is  well suited to describe the system. The crossover
between the two regimes is smooth where the step structure between
the Landau level edges turn into Friedel oscillations. Increasing
temperature smears out the step structure in the two dimensional
regime, and similarly smoothes out the Friedel oscillation in the
one dimensional regime.

The exact wavefunctions given for the rotating anisotropic trap
provide a smooth one to one mapping between the states in a 2D
Landau level and the states in a one dimensional harmonic trap. We
expect this mapping to be useful for study of interacting Bose and
Fermi systems, going beyond the non-interacting gas in this paper.
It would be interesting to see what kind of states are obtained for
an extremely elongated trap if this mapping is used on correlated
states such as the fractional quantum Hall states in the two
dimensional regime. Similarly, certain strongly interacting models
in one dimension may accept continuation into two dimension,
extending the validity of some exact solutions.

Experiments on rotating ultracold gases have so far focused on
isotropic traps, as they mostly rely on the isotropic trap to
conserve angular momentum given to the system before equilibrium is
established. Nevertheless a rotating anisotropic trap has been
demonstrated\cite{Dalibard}. A wide array of experiments have
displayed the feasibility of designing traps that are strongly
confining in one or two dimensions to probe low dimensional physics
of quantum gases. All these successes with trap design lead us to
conclude that a the trap potential discussed in this paper is within
the current experimental capabilities.

A major difficulty in Fermion experiments is creating low enough
temperatures to probe quantum mechanical nature of the
gas\cite{DeMarco, Truscott, Schreck, Granade, Hadzibabic, Roati}.
However as the system considered in this paper is non-interacting,
the temperatures needed to observe qualitative features such as
density steps is well above the currently obtained lowest
temperatures.

The main method employed in measuring density profiles is expansion
imaging of the gas, and this method can be used to determine the
main features of the density profile such as Landau level step
structure or the semicircular density profile in the one dimensional
regime. More subtle features, such as Friedel oscillations would
require more complicated probes of density such as Bragg
spectroscopy\cite{Stenger}.

Finally, we remark that the Fermi gas in a rotating anisotropic trap
is system that may elucidate the connection between one dimensional
physics with dynamics within a Landau level. It would be interesting
to extend the study in this paper to interacting fermion systems.

\section{appendix}

The ground state wave function of the system is written as  \bea
\label{GroundState} \phi_{00}=\frac{1}{\sqrt{\pi a_x a_y}}
\exp{\left[-\frac{x^2}{2 a_x^2} - \frac{y^2}{2 a_y^2} \right]}
\exp{\bigg\{i \frac{M x y}{\hbar} \left[\frac{\gamma}{1+\beta_+
\beta_-} - \frac{1}{2} \left(\frac{\omega_+}{\beta_+} +
\frac{\omega_-}{\beta_-} \right) \right] }\bigg\}, \eea and the
excited states are obtained by applying the relevant raising
operators defined by Fetter\cite{Fetter}on ground state
wavefunction. Operator $\alpha_+$gives the The excited sates wave
functions in each LL, and another one i.e. $\alpha_-$ gives the
higher LL states wavefunctions. The wave functions in LLL are
obtained to be
\bea\label{LLLwavefunctions}\varphi_{m0}(x,y)=\frac{1}{\sqrt{m!}}\varphi_{00}(x,y)P_m(\xi_+)\eea
where \bea
P_m(\xi_+)=\Big(\frac{c}{2}\Big)^{n/2}H_m\bigg(\frac{\xi_+}{\sqrt{2c}}\bigg)
\eea The wavefunctions in higher excited Landau levels also are
found to be \bea
\varphi_{m1}(x,y)=\frac{1}{\sqrt{m!}}\varphi_{00}(x,y)\bigg[\xi_-P_m(\xi_+)-2im\rho
P_{m-1}(\xi_+)\bigg], \eea \bea \varphi_{m2}(x,y)&=&
\frac{1}{\sqrt{2m!}}\varphi_{00}(x,y)\bigg\{\Big(\xi_-^2-c\Big)P_m(\xi_+)\nonumber\\
&-&4im\rho\xi_-P_{m-1}(\xi_+)\nonumber\\&-&4m(m-1)\rho^2P_{m-2}(\xi_+)\bigg\},\eea
\bea\varphi_{m3}(x,y)&=&
\frac{1}{\sqrt{3!m!}}\varphi_{00}(x,y)\bigg\{\Big(\xi_-^3-3c\xi_-\Big)P_m(\xi_+)\nonumber\\&-&6im\rho\Big(\xi_-^2-c\Big)P_{m-1}(\xi_+)\nonumber\\
&-&12m(m-1)\rho^2\xi_-P_{m-2}(\xi_+)\nonumber\\&+&8im(m-1)(m-2)\rho^3P_{m-3}(\xi_+)\bigg\}\eea
\bea \varphi_{m4}(x,y)&=&
\frac{1}{\sqrt{24m!}}\varphi_{00}(x,y)\bigg\{\Big(\xi_-^4-6c\xi_-^2+3c^2\Big)P_m(\xi_+)\nonumber\\&-&8mi\rho\Big(\xi_-^3-3c\xi_-\Big)P_{m-1}(\xi_+)\nonumber\\
&-&24m(m-1)\rho^2\Big(\xi_-^2-c\Big)P_{m-2}(\xi_+)\nonumber\\&+&32im(m-1)(m-2)\rho^3\xi_-P_{m-3}(\xi_+)\nonumber\\&+&16m(m-1)(m-2)(m-3)\rho^4P_{m-4}(\xi_+)\bigg\}\eea
\bea\varphi_{m5}(x,y)&=&
\frac{1}{\sqrt{120m!}}\varphi_{00}(x,y)\bigg\{\Big(\xi_-^5-10c\xi_-^3+15c^2\xi_-\Big)P_m(\xi_+)\nonumber\\&-&10im\rho\Big(\xi_-^4-6c\xi_-^2+3c^2\Big)P_{m-1}(\xi_+)\nonumber\\
&-&40m(m-1)\rho^2\Big(\xi_-^3-3c\xi_-\Big)P_{m-2}(\xi_+)\nonumber\\&+&80im(m-1)(m-2)\rho^3\Big(\xi_-^2-c\Big)P_{m-3}(\xi_+)\nonumber\\&+&80m(m-1)(m-2)(m-3)\rho^4\xi_-P_{m-4}(\xi_+)\nonumber\\&-&32im(m-1)(m-2)(m-3)(m-4)\rho^5P_{m-5}(\xi_+)\bigg\}\eea

We suggest a general form for the wavefunctions of the excited
states: \bea
\varphi_{mn}(x,y)&=&\frac{1}{\sqrt{n!m!}}\varphi_{00}(x,y)\nonumber\\
&\times&\sum_{k=0}^n\bigg[
(-i)^{n-k}\frac{2^{n-k}}{(n-k)!}\rho^{n-k}\nonumber\\
&\times&\frac{d^{n-k}}{d^{n-k}\xi_-}P_n(\xi_-)
\frac{d^{n-k}}{d^{n-k}\xi_+}P_m(\xi_+) \bigg]\eea

\acknowledgements

N.G. is supported by T\"{U}B\.{I}TAK. M. \"{O}. O. is supported by
T\"{U}B\.{I}TAK-KAR\.{I}YER Grant No. 104T165.


\begin{thebibliography}{99}


\bibitem{Dalibard} K. W. Madison, F. Chevy, W. Wohlleben, J.
Dalibard, Phys. Rev. Lett. {\bf 84}, 806, (2000).
\bibitem{Ketterle}J. R.
Abo-Shaeer, C. Raman, J. M. vogels, and W. Ketterle, Science {\bf
292}, 476 (2001).
\bibitem{Cornell}P. C. Haljan, I. Coddington, P.
Engels, E. A. Cornell, Phys. Rev. Lett. {\bf 87}, 210403 (2001).
\bibitem{Cooper}N.R. Cooper, N.K. Wilkin, and J.M.F. Gunn, Phys. Rev. Lett.
{\bf 87}, 120405 (2001).
\bibitem{Sinova}J. Sinova, C.B. Hanna, and A.H. MacDonald, Phys.
Rev. Lett. {\bf 89}, 030403 (2002).
\bibitem{Baranov} M. A. Baranov, K. Osterloh, and M. Lewenstein, Phys. Rev. Lett.
{\bf 94}, 070404 (2005).
\bibitem{Sorensen} A. S. Sorensen, E. Demler, and M. D.
Lukin, Phys. Rev. Lett. {\bf 94}, 086803 (2005).
\bibitem{Hafezi} M. Hafezi, A. S. Sorensen, E. Demler, and M. D.
Lukin, Phys. Rev. A {\bf 76}, 023613 (2007).
\bibitem{Osterloh} K. Osterloh, N. Barberan, and M. Lewenstein, Phys. Rev. Lett. {\bf 99}, 160403 (2007).
\bibitem{Greiner} M. Geriner, C. Regal, D. Jin, Nature {\bf 426}, 537
(2003).
\bibitem{Zwierlein} M. W. Zwierlein, C. A. Stan, C. H. Schunck, S. M.
F. Raupach, A. J. Kerman, W. Ketterle, Phys. Rev. Lett. {\bf 92},
120403 (2004).
\bibitem{Zwierlein2} M. W. Zwierlein, J. Abo-Shaeer, A. Shirotzek, C.
H. Schunch, W. Ketterle, Nature {\bf 435}, 1047 (2005).
\bibitem{Bloch} I. Bloch, J. Daliabrd, W. Zwerger, Rev. Mod. Phys.
{\bf 80}, 885 (2008).
\bibitem{Dalibard2} P. Rosenbusch, D. S. Petrov, S. Sinha, F. Chevy, V. Bretin, Y. Castin, G. Shlyapnikov, and J.
Dalibard, Phys. Rev. Lett. {\bf 88}, 250403 (2002).
\bibitem{linn} M. Linn, M. Niemeyer, A. L. Fetter, Phys. Rev. A
{\bf 64}, 023602 (2001).
\bibitem{Oktel} M. ~\"O.~ Oktel, Phys. Rev. A {\bf 69}, 023618
(2004).
\bibitem{Fetter} A. L. Fetter, Phys. Rev. A {\bf 75},  013620
(2007).
\bibitem{Sinha}S. Sinha, G. V. Shlyapnikov, Phys. Rev. Lett. {\bf
94}, 150401 (2005).
\bibitem{Sanchez} P. Sanchez-Lotero, J.J. Palacios, Phys. Rev. A {\bf
72}, 043613 (2005).
\bibitem{Aftalion} A. Aftalion, X. Blanc, N. Lerner, Phys. Rev A {\bf 79},
011603(R) (2009).
\bibitem{Ho} T. L. Ho, C. V. Ciobanu, Phys. Rev. Lett. {\bf 85}, 4648 - 4651
(2000).
\bibitem{Arfken}G. B. Arfken, H. J. Weber, Mathematical Methods for
Physicists (Elsvier Academic Press, 2005).
\bibitem{Freidel} J. Friedel, Nuovo Cimento (Suppl.) 7, {\bf287}
(1958).
\bibitem{Abramowitz}M. Abramowitz and I. Stegun, Handbook of Mathematical Functions,
(Dover, New York, 1970).
\bibitem{DeMarco} B. DeMarco, and D. S. Jin, Science {\bf 285}, 1703 (1999).
\bibitem{Truscott} A. G. Truscott, K. E. Strecker, W. I. McAlexander, G. B.
Partridge, and R. G. Hulet, Science {\bf 291}, 2570 (2001).
\bibitem{Schreck} F. Schreck, L. Khaykovich, K. L. Corwin, G. Ferrari,
T. Bourdel, J. Cubizolles, and C. Salomon, Phys. Rev. Lett. {\bf
87}, 080403 (2001).
\bibitem{Granade} S. R. Granade, M.E. Gehm, K. M. O'Hara, and J. E. Thomas,
Phys. Rev. Lett. {\bf 88}, 120405 (2002).
\bibitem{Hadzibabic} Z. Hadzibabic, C. A. Stan, K. Dieckmann, S. Gupta, M.W. Zwierlein, A. Gorlitz, and W.
Ketterle, Phys. Rev. Lett. {\bf 88}, 160401 (2002).
\bibitem{Roati} G. Roati, F. Riboli, G. Modugno, and M. Inguscio, Phys. Rev. Lett. {\bf
89}, 150403 (2002).
\bibitem{Stenger} J. Stenger, S. Inouye, A. P. Chikkatur, D. M. Stamper-Kurn, D. E. Pritchard, and W. Ketterle, Phys. Rev. Lett. {\bf 82}, 4569-4573 (1999).


\end{thebibliography}
\end{document}